\documentclass[preprint, 12pt]{elsarticle}

\usepackage{amsmath,amssymb,amsfonts}
\usepackage[utf8]{inputenc}
\usepackage[english]{babel}
\usepackage{tabularx, multirow}
\usepackage{algorithmic}
\usepackage{textcomp}
\usepackage{xcolor}
\usepackage{fancyhdr,lipsum}
\usepackage[colorlinks=true,]{hyperref}
\usepackage[official]{eurosym}
\usepackage{todonotes}
\usepackage{bookmark} %
\usepackage{placeins}
\usepackage{enumitem}
\usepackage{float}

\begin{document}
\emergencystretch 3em

\begin{frontmatter}
    \title{Offshore power and hydrogen networks for Europe’s North Sea}
    \author{Philipp Glaum}
    \ead{p.glaum@tu-berlin.de}
    \author{Fabian Neumann}
    \author{Tom Brown}
    \address{Technische Universität Berlin, Department of Digital Transformation in Energy Systems, Institute of Energy Technology, Berlin 10623, Germany}
    \begin{abstract}
        The European North Sea has a vast renewable energy potential and can be a powerhouse for Europe's energy transition.
        However, currently there is uncertainty about how much offshore wind energy can be integrated, whether offshore grids should be meshed and to what extent offshore hydrogen should play a role.
        To address these questions, we use the open-source energy system optimization model PyPSA-Eur to model a European carbon-neutral sector-coupled energy system in high spatial and temporal resolution.
        We let the model endogenously decide how much offshore wind is deployed and which infrastructure is used to integrate the offshore wind.
        We find that with point-to-point connections like we have today, 310~GW offshore wind can be integrated in the North Sea.
        However, if we allow meshed networks and hydrogen, we find that this can be raised to 420~GW with cost savings up to 15~bn\euro/a.
        Furthermore, we only observe significant amounts of up to 75~GW of floating wind turbines in the North Sea if we have offshore hydrogen production.
        Generally, the model opts for offshore wind integration through a mix of both electricity and hydrogen infrastructure.
        However, the bulk of the offshore energy is transported as hydrogen, which is twice as much as the amount transported as electricity.
        Moreover, we find that the offshore power network is mainly used for offshore wind integration, with only a small portion used for inter-country transmission.
    \end{abstract}
\end{frontmatter}

\pagebreak
\section{Introduction}

The European Union (EU) has set a goal to achieve climate neutrality by 2050, a target outlined in the European Green Deal of 2020~\cite{europeanparliamentEuropeanGreenDeal2020}.
To realize this ambitious objective, a significant expansion of renewable energy sources is necessary.
At the same time there is a projected increase in the total energy demand for electricity and hydrogen within the European single market~\cite{europeancommission.directorategeneralforenergy.METISStudyS52023}.
The strong rise in electricity demand originates from other sectors electrifying.
On the other hand, hydrogen is particularly relevant for sectors that are challenging to electrify.
For shipping and aviation as well as for feedstocks in the chemicals industry and for steelmaking, hydrogen can be used or transformed to create methane, methanol, ammonia or Fischer-Tropsch fuels.

To serve this increasing demand, many view the European North Sea (NS) with its shallow waters as a promising area for offshore wind energy, potentially evolving into an energy hub.
For this reason, the governments of the NS countries have committed to establishing at least 300~GW of offshore wind capacities in the NS by 2050~\cite{OstendDeclarationNorth2023}.
Wind energy is essential in this context for several reasons: its seasonal alignment with demand, its abundant availability in Europe, its cost-effectiveness, its contribution to strategic autonomy and energy security, and the presence of a competitive EU wind industry~\cite{europeancommissionEUWindPower2023}.
Especially for green hydrogen and derivatives, offshore wind can play an important role~\cite{muellerWindtoHydrogenTechGoes2023}.
With an estimated cost-effective wind potential of 635~GW~\cite{europeancommission.directorategeneralforenergy.HybridProjectsHow2019} by 2030, the NS could provide a substantial portion of Europe’s energy demand.
This potential is not only vast but is also advantageously located close to high-demand areas with a population of 200 million and 20\% of Europe's GDP~\cite{martinez-gordonModellingHighlyDecarbonised2022}.

The slow expansion of onshore wind energy, often delayed by public acceptance issues~\cite{hevia-kochComparingOffshoreOnshore2019}, further underscores the importance of offshore wind energy, which faces fewer acceptance problems~\cite{linnerudPeoplePreferOffshore2022a}.
This is also illustrated by the notable gap between the estimated technical onshore wind potential of 13.4~TW~\cite{rybergFutureEuropeanOnshore2019} in Europe and the 1~TW anticipated for deployment by the European wind industry by 2050~\cite{novakWindWillBe2021a}.

To fully harness the wind potential of the NS, one needs to consider the deployment of floating wind.
Floating wind turbines allow for the exploitation of wind resources in deeper waters with comparable high capacity factors, representing the next frontier in the offshore wind industry~\cite{maienzaLifeCycleCost2020}.
There are already operating pilot projects for floating wind demonstrating its feasibility for example in Scotland and Spain~\cite{equinorHywindScotland,saitecDemoSATHStartsGeneration2023}.
Although commercial floating wind installations are still uncommon in Europe, the market is rapidly evolving, with numerous projects expected to be operational by 2030~\cite{diazMarketNeedsOpportunities2022}.
In the UK for example, shallow water seabeds are already fully allocated for fixed-bottom turbines.
Therefore, floating wind emerges as the sole option to exploit the remaining offshore seabeds.
Consequently, the UK has initiated auctions for seabed leases specifically designated for floating wind projects~\cite{fairleyGlobalRaceTap2023}.

\begin{figure}[!htt]
    \centering
    \includegraphics[width=.7\textwidth]{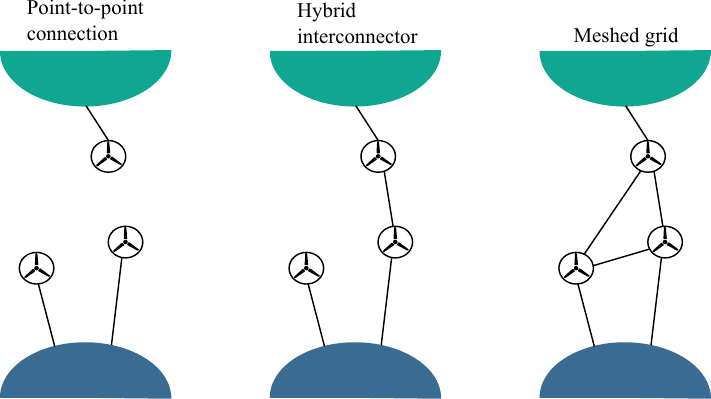}
    \caption{Schematic illustration of the different offshore connection types. The traditional point-to-point connection connecting the wind farm directly to shore, the hybrid interconnector connecting two countries while integrating wind energy and the meshed offshore grid connecting multiple countries and wind farms.}
    \label{fig:offshore_topologies}
\end{figure}

Integrating the wind energy from the NS into the onshore energy system poses significant challenges, since it requires the development of new long-distance transmission infrastructure.
Transmission System Operators (TSOs) have already expressed interest in building such infrastructure in the North Sea~\cite{northseawindpowerhubconsortiumHubsSpokesViable2022}.
Traditionally, offshore transmission lines have a national focus, connecting wind farms directly to the shore via radial connections, also known as point-to-point connections~\cite{martinez-gordonBenefitsIntegratedPower2022a}.
However, new hybrid interconnectors aim to connect not only the wind farms but also facilitate inter-country transmission between two countries~\cite{entso-eENTSOEPositionOffshore2020}.
These hybrid structures align with the common interests of NS countries~\cite{OstendDeclarationNorth2023} and the European Commission~\cite{europeancommissionProjectsCommonInterest2022}.
A further extension of hybrid interconnectors would be the deployment of a more meshed offshore grid, not only connecting two countries but having terminals connecting to multiple countries.
Fig.~\ref{fig:offshore_topologies} illustrates the different offshore connection types.

While hybrid projects offer numerous benefits, concerns about the legal and regulatory barriers have been voiced like uncertainty of project responsibility~\cite{europeancommission.directorategeneralforenergy.HybridProjectsHow2019}.
However, ENTSO-E as association of European TSOs highlights that an integrated offshore grid can improve energy security, enhance trading opportunities, drive standardization, and reduce the need for overall assets~\cite{entso-eOffshoreGridDevelopment2011}.

One example for an offshore infrastructure project is the North Sea Wind Power Hub~\cite{northseawindpowerhubconsortiumHubsSpokesViable2022}.
Behind the project is a consortium of TSOs and gas TSOs, which aims to construct a substantial offshore wind hub in the NS, connecting it to the onshore grid via High Voltage Direct Current (HVDC) cables.
Besides electric interconnectors the hub is also envisioned to have hydrogen production facilities on the hubs and pipelines connecting to onshore.

Looking closer at offshore hydrogen production, the H2Mare project wants to investigate the cost-effectiveness of offshore versus onshore electrolysis.
It explores various concepts, such as producing hydrogen offshore and transporting it onshore through pipelines, transferring electricity onshore for electrolysis, or producing and storing hydrogen offshore in floating tanks for collection by ships~\cite{muellerWindtoHydrogenTechGoes2023}.

These industry projects are underscored by academic research on how to integrate offshore wind into the energy system.
With several approaches using mainly expansion and dispatch optimization problems, researchers have evaluated the benefits of an offshore network.
In~\cite{gorensteindedeccaExpansionPlanningNorth2018} and~\cite{martinez-gordonModellingHighlyDecarbonised2022}, the authors investigate an offshore power grid for the integration of specific offshore wind locations.
As well as the offshore power grid, the authors of~\cite{gea-bermudezGoingOffshoreNot2023},~\cite{durakovicPoweringEuropeNorth2023a},~\cite{luthHowConnectEnergy2023},~\cite{martinez-gordonBenefitsIntegratedPower2022a} also consider offshore hydrogen infrastructure with electrolysis and pipelines.
The authors of the studies generally are aligned in their assessment that an offshore grid increases system welfare and leads to better integration of offshore wind.
Furthermore, they agree that additional offshore hydrogen production leads to reduced curtailment of offshore wind as well as system cost reduction.
However, the authors' conclusions diverge regarding the extent of the cost benefit of offshore hydrogen production.
While~\cite{gea-bermudezGoingOffshoreNot2023} say that offshore hydrogen production yields only small system benefits of 1~bn\euro/a,~\cite{martinez-gordonBenefitsIntegratedPower2022a} finds that offshore hydrogen production complements onshore production leading to benefits of 4~bn\euro/a.
Most studies only consider NS countries in their modelling and often do not include other sectors besides electricity and hydrogen.
For this reason, they cannot reflect the dynamics in the European energy system and the interplay between different sectors which become more important in a carbon-neutral future.
Furthermore, their spatial resolution in the NS is often limited to a few offshore regions, up to a maximum of 15 in~\cite{durakovicPoweringEuropeNorth2023a}, and does not offer a broader set of offshore topology options.
To rigorously model different offshore topologies, we argue that a higher spatial resolution in the NS is essential.

The novelty of our paper is simultaneously evaluating more different offshore network options in high resolution while also considering all energy and non-energy sectors including electricity, industry, transport, agriculture and households.
The inclusion of diverse sectors is crucial to accurately assess the energy demand of the different sector, which could exhaust local potentials, and to capture complex dynamics between the sectors.
Moreover, the model must incorporate a detailed allocation of offshore resources and apply realistic cost assumptions for both wind turbines and offshore connection costs, including the consideration of wake effects.
All these features are combined in our model, to let the model endogenously decide the optimal offshore infrastructure for the NS.
Besides a more extensive model, our study provides a sensitivity analysis testing our results for robustness against different uncertainties like the feasibility of large-scale onshore wind or transmission expansion.
With our study, we aim to determine the optimal characteristics and system benefits of offshore networks, analyze the trade-offs between electricity and hydrogen transmission in the North Sea, assess the role of floating wind in various offshore topologies, and evaluate whether the offshore grid is primarily used for wind integration or also for inter-country energy transport.

\section{Methodology}
\label{sec:methodology}
\subsection{Energy System Model}
\label{sec:pypsa-eur}

In our study, we employ the open-source model PyPSA-Eur~\cite{horschPyPSAEurOpenOptimisation2018} to optimize the European energy system.
PyPSA-Eur deploys the framework PyPSA~\cite{brownPyPSAPythonPower2018}, an open-source tool,  and fills it with data to represent the European energy system.
To incorporate the data, PyPSA-Eur uses the workflow manager Snakemake~\cite{molderSustainableDataAnalysis2021}, which ensures reproducibility and handles the preprocessing in the model.

PyPSA-Eur is a graph-based model, meaning that the energy system is represented as a graph with nodes and edges connecting the nodes.
The nodes represent injection and extraction points, connecting to different components for injection like generators, for extraction like loads, or both e.g.~storages.
On the other hand, the edges represent the energy flows between the nodes, e.g.~transmission lines or gas pipelines.
A key feature of our model is its sector-coupled approach, which integrates multiple sectors including electricity, transport, industry, agriculture and heat.
The optimization model behind PyPSA-Eur is linear, aiming to minimize both investment and operational costs of the energy system.
In the model, we optimize different technologies like wind, solar, heat pumps, power-to-X and network infrastructure for a single year.
Furthermore, we consider in total 33 European countries.
The optimization model takes into account various essential constraints, such as nodal energy balance, system CO2 emission limits, and grid constraints.
For example in the grid constraint, we consider physical correlations like the Kirchhoff voltage law for transmission lines.

Relevant data sources in the model are the existing grid infrastructure, the energy demands in the regarded sectors and the techno-economic data.
The grid infrastructure data is taken from ENTSO-E~\cite{entso-eGridMap}, which is extended by several TYNDP projects from~\cite{entso-ePlanningFutureGrid}.
Most energy demands for the different sectors and countries are obtained from the "Integrated Database of the European Energy System" of the Joint Research Center~\cite{jointresearchcenterIntegratedDatabaseEuropean2015}.
For countries outside the EU, we retrieve the energy balances from national datasets.
The countrywide annual demands are then distributed with heuristics to resolve the demands regionally.
For the techno-economic data, we use a technology dataset~\cite{aarhusuniversityPyPSATechnologyData2023} that contains data on the technologies used in the model.
The technology data is mainly sourced from the Technology Catalog of the Danish Energy Agency\cite{danishenergyagencyCostPerfomanceData2023}.
Additionally, to obtain weather-dependent data such as renewable resource data, residential heating demands, coefficients of performance for heat pumps, we use the Atlite toolbox~\cite{hofmannAtliteLightweightPython2021}.
For example for renewable resources, Atlite converts weather data, obtained from ERA5~\cite{copernicusclimatechangeservicec3sERA5HourlyData2023} reanalysis data and SARAH-2 data~\cite{pfeifrothSurfaceRadiationData2017}, into renewable energy potentials and time series.
Furthermore, for renewable resources, we also consider exclusion areas in up to 100$\times$100m resolution, such as nature reserves, other land usages and shipping lanes.
Shipping lanes are for instance sourced as raster data from the Global Shipping Traffic Density~\cite{theworldbankGlobalShippingTraffic2020}, tracking hourly signals from different vessel types like commercial or fishing ships.
\subsection{Offshore Modelling}\label{sec:offhore-modelling}

To enhance the modelling of offshore wind resources in the PyPSA-Eur model, we focus on improving five aspects of our offshore modelling, described in the following sections.

\subsubsection*{1. Increase offshore resolution}
The land eligibility calculation with Atlite results in a location-specific renewable resource and total installation potential per model region.
In PyPSA-Eur a model region represents a geographical area or shape attributed to a specific node.
For the offshore regions, each country with offshore shapes in its exclusive economic zone has at least one offshore region.
The number of nodes we attribute to the offshore shape determine the number and size of the offshore regions in each country.
We use a Voronoi diagram to allocate the area of the offshore shapes to the regions.
The Voronoi diagram divides the offshore shape into regions, where each region is closer to its node than to any other node.
If a country has only a few offshore nodes while having a large offshore shape, the offshore regions will also be large elongated regions.
Since every offshore region only has a single time series and potential, this leads to inaccurate modeling of the offshore region for example for the capacity factors which vary spatially.

\begin{figure}[h!]
    \centering
    \includegraphics[width=.6\textwidth]{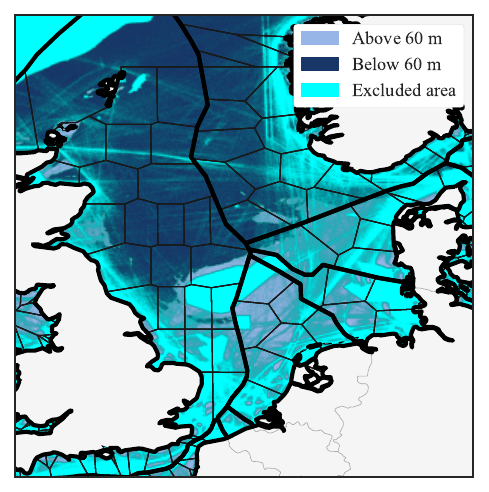}
    \caption{Available area for offshore wind in the North Sea for water depths above and below 60~m.}
    \label{fig:available_area}
\end{figure}

To address this, we increase the number of offshore nodes in the model, resulting in smaller offshore regions.
In Atlite the highest spatial resolution for geographical dependent potentials and time series is 30$\times$30km, determined by the resolution of the ERA5 raster cell.
However, in the standard PyPSA-Eur model, the size of offshore regions significantly exceeds the 30$\times$30km raster cell size of the weather data.
As a consequence, Atlite aggregates all raster cell data which lies in the offshore region to obtain a single times series and potential for the region.
In response to this, we divide the offshore regions into smaller segments, with a maximum size of approximately 10,000~km$^2$.
This maximum size derives from the trade off between detail and computation time.
The increase in offshore regions allow for a more precise allocation of resources to each offshore region, preventing the aggregation of resources from different geographical locations.
Furthermore, we have modeled three distinct offshore wind technologies varying in costs within these regions: near and far fixed bottom, and floating wind turbines.
The offshore wind technologies distinguish themselves based on their maximum allowable water depth and distance from the shore as shown in Tab.~\ref{tab:offshore_technologies}.
\begin{table}[!htt]
    \caption{Maximal water depth and permissive distances from shore for the different offshore wind technologies.}
    \label{tab:offshore_technologies}
    \begin{tabular*}{\textwidth}{l|r|r|r}
        \hline
        \textbf{Technology} & \textbf{Max. Depth} & \textbf{Min. Distance}& \textbf{Max. Distance} \\
        \hline
        Near Fixed Bottom & 60 m & 1 km & 30 km  \\
        Far Fixed Bottom & 60 m & 30 km & - \\
        Floating & 1000 m & 1 km & - \\
        \hline
    \end{tabular*}

\end{table}
Each offshore technology is then assigned a specific potential and capacity factor based on its available area.
Fig.~\ref{fig:available_area} illustrates the available area in the NS for the water depths below 60~m available for fixed bottom, and above 60~m for floating wind installations.

\subsubsection*{2. Turbine cost modelling}

\begin{figure}[!ht]
    \centering
    \includegraphics[width=0.7\textwidth]{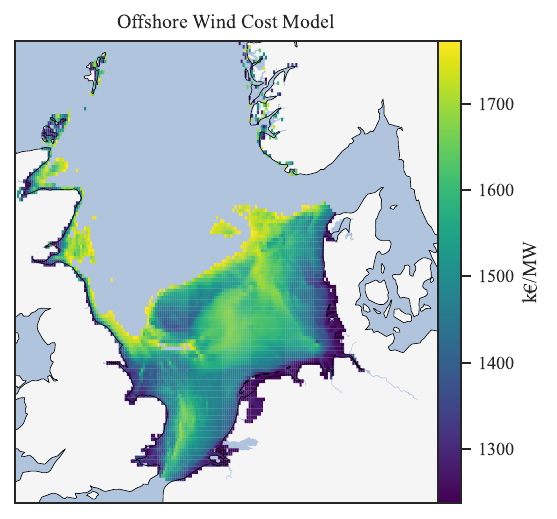}
    \caption{Location specific cost in the North Sea for a 12~MW fixed bottom wind turbines according to cost model from Danish Energy Agency. The wind turbine has a hub height of 136~m and a rotor diameter of 214~m.}
    \label{fig:offwind_cost}
\end{figure}

To enhance the cost representation of wind turbines, we have incorporated the wind turbine cost model from the Danish Energy Agency~\cite{jorgensen21WindTurbines}.
This model estimates costs for fixed bottom offshore wind turbines while considering technical data such as hub height, rotor diameter, specific power, inter-array distance, and water depth.
With these parameters, the model computes the investment and installation costs for the turbine, its foundation, and the inter-array cabling.
The site specific turbine cost for our 12~MW reference offshore wind turbine from~\cite{nrelNRELTurbineModels} is shown in Fig.~\ref{fig:offwind_cost}.
In our model, the costs for the wind turbines do not include the cost for the grid connection as these are endogenously optimized.
Due to the different water depths, the investments cost for fixed bottom turbines vary between 1200~k\euro{}/MW and 1800~k\euro{}/MW.
In the North Sea, one cannot generally say that offshore wind turbine cost increase with distance to shore due to the strong variation of the water depth.
For example the Dogger Bank has comparable low turbine costs despite its distance to shore.

Contrary to fixed bottom, floating wind turbines costs are assumed to be constant throughout different locations with 2100~k\euro{}/MW, following~\cite{mckennaAnalysingLongtermOpportunities2021}.
In line with Sykes et al.~\cite{sykesReviewAnalysisUncertainty2023}, we ignore the depth-dependent costs for mooring lines of floating wind turbines.

\subsubsection*{3. Wake effect modelling}

In line with the approach adopted by Gea-Bermúdez et al.~\cite{gea-bermudezInfluenceLargeScaleWind2021}, we have integrated wake effect modelling into our model.
This addition takes into account the wake losses that occur as more offshore wind capacities are installed in the same region.

We have structured the wake loss rates based on installed capacity thresholds.
For the initial 2~GW of installed capacity per region, we assume wake losses of 9\% reduction in capacity factor.
As the capacity increases by another 10~GW, we estimate the wake losses for this segment at 21\%, and for any additional capacity beyond this, the losses are projected to be 32\%.

The high wake loss rates for capacities exceeding 2 GW are to account for the added wake loss effects on the first 2 GW if additional capacities are built.
In this way, we obtain appropriate wake effects for the whole wind power fleet.

\subsubsection*{4. Offshore power network}

Each offshore region has the option to build an offshore HVDC platform, which converts the AC power from the wind farms to DC power.
These offshore platforms encompass the necessary substructures and converter stations required for a HVDC system, with cost data derived from Vrana et al.~\cite{vranaImprovedInvestmentCost2023}.
The assumed cost in our model for the offshore HVDC platform is 480~\euro/kW$_{el}$ while for an onshore HVDC converter station it is 120~\euro/kW$_{el}$.

\begin{figure}[!ht]
    \centering
    \includegraphics[width=0.6\linewidth]{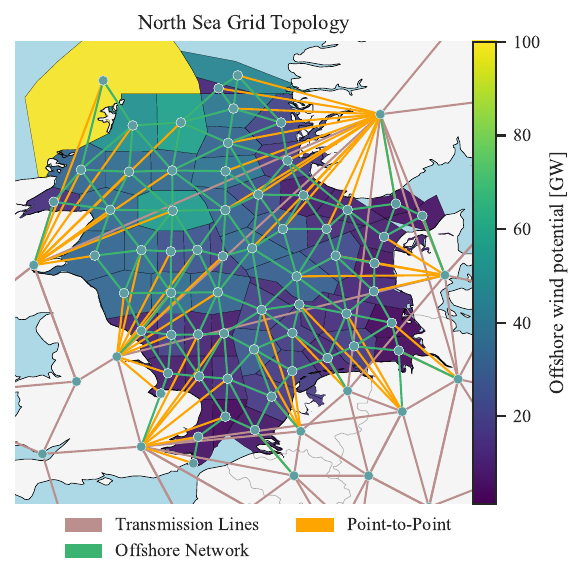}
    \caption{Possible transmission candidates of offshore grid in the North Sea. The topology is the same for the power grid and the hydrogen gird.}
    \label{fig:offshore_network}
\end{figure}

The offshore regions can be interconnected via HVDC cables facilitating energy transfer between them, costing 0.9~\euro/kW$_{el}$/km.
In addition to interconnecting offshore regions, these regions are also linked through HVDC cables to the closest onshore region of the same country representing radial connections.
For some cases, there is a closer onshore region in a neighboring country.
In this case, we connect the offshore region to both the closest overall onshore region and the closest onshore region in the same country.
This setup allows the model to endogenously determine the connectivity of offshore regions, deciding whether they are interconnected, radially connected to onshore regions, or a combination of both.
We demonstrate the offshore power network topology available to the model in Fig.~\ref{fig:offshore_network}.
Besides the topology, we highlight the offshore wind potentials of the different regions.
We model the flow of power in the meshed HVDC cables using a transport model where the flow in each line is controllable.
We do not include the Kirchhoff Voltage Law because for building brand-new networks this requires a mixed-integer linear modelling approach, which would be intractable for problems of our size.

For offshore cabling in the NS, we exclusively use HVDC cables, as the distance from the centroid of the offshore regions to onshore regions always exceeds 80~km, making Alternating Current (AC) cabling impractical.
Hence, in the NS all offshore wind farms, including near-shore, can only connect to the onshore network via HVDC links.
This approach is a simplification adopted for our modeling of the offshore network.
Near-shore wind farms outside the NS are connected to the onshore network via AC cables.

For the AC transmission system, we consider a piecewise-linear power loss approximation outlined in~\cite{neumannApproximatingPowerFlow2020}.
In contrast, for DC transmission system, we apply for every DC line a constant loss rate of 2\% and a dynamic loss rate of 2.3\% per 1000~km completely neglecting non-linear behaviors.
The non-linear behavior for DC derives from the higher losses per power transfer with higher loading.

\subsubsection*{5. Offshore hydrogen network}

In our model, the possible topology of the offshore hydrogen network mirrors the offshore power network.
Consequently, each offshore region is also equipped with a corresponding offshore hydrogen platform.
Therefore, hydrogen has the same eligible offshore network topology as HVDC (Fig.~\ref{fig:offshore_network}), just with hydrogen platforms and pipelines instead of HVDC infrastructure.
The hydrogen platforms include key components like the electrolysis, substructure, and a desalination unit for the water supply of the electrolysis.
They cost 440~\euro/kW$_{el}$ sourced from the reference case of a study by the Danish Energy Agency~\cite{danishenergyagencyCostPerfomanceData2023}.
Additionally, the submarine hydrogen pipelines are costed at 0.5~\euro/kW$_{H2}$/km.

Besides their electrical connection, each offshore wind farm is connected to an offshore hydrogen platform.
This dual connectivity enables the model to endogenously determine the proportion of offshore wind energy allocated for hydrogen production versus the amount fed into the HVDC grid.

Similar to the HVDC cables, the model accounts for losses of the hydrogen pipelines.
These losses are calculated at 1.9\% per 1000~km, reflecting the electricity consumption for both compression and transport~\cite{enagasEuropeanHydrogenBackbone2020}.

\section{Study Case}
\label{sec:study-case}

For our study, we model the European energy system considering 33 countries with a carbon-neutrality constraint using techno-economic projections for the year 2030.
To see the modelled countries and selected cost assumptions see Table~\ref{tab:model_countries} and\ref{tab:cost_assumptions}.
The spatial representation in our model includes 130 regions, 64 onshore regions and 66 offshore regions.
We set the temporal resolution in the model to a 3-hourly resolution for one year.
In this way we have enough detail to capture the variability of renewable energy resources and storage operation.
For weather data we take the weather year of 2013 which is the default year in PyPSA-Eur.
Our model is partially greenfield which means there are only existing capacities for transmission lines, hydropower, wind and solar.

For our offshore network modelling, we vary four parameters:
\begin{itemize}
    \item \textbf{P2P vs. meshed power network:} We differentiate between models that incorporate a meshed offshore power network and those limited to point-to-point (P2P) connections, which link offshore parks directly to onshore landing points.
    \item \textbf{W/o vs. P2P  vs. meshed H2 network:} Certain models allow offshore hydrogen production and transport, whereas others are limited to onshore hydrogen production only.
    In the model we allow hydrogen transport either through P2P connections or a meshed offshore hydrogen network.
    \item \textbf{Onshore transmission:}  We examine various models regarding the expansion of onshore transmission capacities.
    While submarine cable expansion is always permitted, restrictions apply only to onshore transmission since this is where most public opposition to new infrastructure can be expected.
    \item \textbf{Onshore wind potential:} The potential for onshore wind energy capacity is varied across different models to test the impact on offshore energy infrastructure. We consider onshore wind potentials of 2.2~TW, 4.4~TW and 8.8~TW, representing 25,50 and 100\% of the technical potential in the considered countries.
\end{itemize}

By varying the given parameters, we construct four main scenarios with different network topologies:
\begin{enumerate}[label=Scenario \arabic*:]
    \item with P2P power network without hydrogen network (reference scenario)
    \item with meshed power network without hydrogen network
    \item with P2P power and hydrogen network
    \item with meshed power and hydrogen network
\end{enumerate} 
We refer to scenario 1 as the reference scenario.
In all of 4 scenarios, we allow for cost-optimal expansion of onshore transmission lines.
However, even if we do not limit the total transmission expansion, there is always an expansion limit of 20~GW per line.
Furthermore, we limit the onshore wind potential to 2200~GW, because we assume that a higher deployment is rather unlikely.

\section{Results}
\label{sec:results}
 First, in Section~\ref{sec:results_offshore_network_comparison}, we make a comparative assessment of the four scenarios with their different offshore network topologies. 
 Following this, we take a closer look at the network topology and capacity allocation of the scenario with the highest cost benefits in Section \ref{sec:results_full_network_scenario} and compare it with the reference scenario.
 Finally, in Section~\ref{sec:results_sensitivity}, we perform a sensitivity analysis of the results to onshore wind potential and onshore transmission capacities.

\subsection{Scenario comparison}
\label{sec:results_offshore_network_comparison}

This section provides a comparative analysis of the reference scenario only with P2P power network without hydrogen network against the three other scenarios.
For the comparison, we keep the parameters for the onshore wind potential at 2.2~TW and the onshore transmission expansion optimal across all scenarios.

\begin{figure}[h!]
    \centering
    \includegraphics[width=1\textwidth]{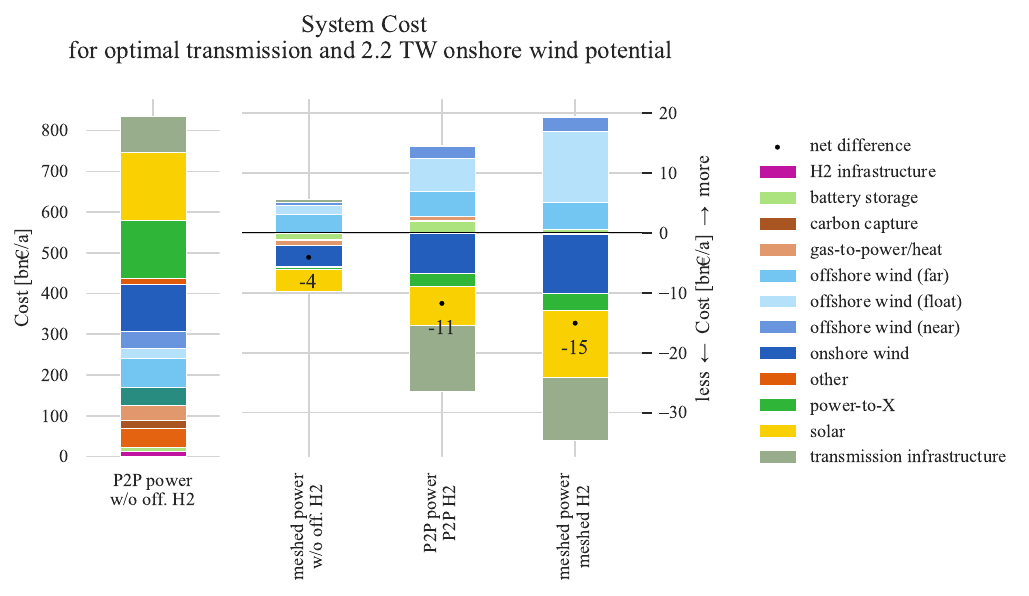}
    \caption{Cost comparison of the four scenarios with their different offshore network topologies. The first column shows the reference case. In the other columns we calculate the cost differences compared to the reference case.}
    \label{fig:ref_cost_comparison}
\end{figure}

Fig.~\ref{fig:ref_cost_comparison} illustrates the cost differences among these topologies.
First, the most left column demonstrates that the total system cost of the reference scenario is approximately 800~bn~\euro{}/a.
Introducing the offshore HVDC network leads to a total system cost benefit of 4~bn~\euro{}/a.
This benefit is attributed to the enhanced integration of cost-effective offshore wind capacities and strengthened intercountry transmission capabilities.
However, since intercountry transmission volumes constitute only about 5\% of the total offshore wind generation, this benefit is relatively modest, see exports in Fig.~\ref{fig:ref_tranmission_comparison}.
Hence, the offshore grid mainly serves to integrate offshore wind capacities and not to transport electricity between countries.

The comparison between the reference scenario and scenarios incorporating hydrogen infrastructure reveals a more significant advantage originating from the hydrogen network rather than from the HVDC network.
This is due to the large demand for hydrogen and derivative products, and the relative cost-efficiency of offshore hydrogen production over constructing HVDC infrastructure for onshore hydrogen production.
Both hydrogen scenarios result in cost savings for offshore transmission infrastructure while integrating significant amounts of offshore wind capacities.

The cost benefit of the scenario with P2P power and hydrogen network amounts to 11~bn~\euro{}/a, which increases to 15~bn~\euro{}/a when additionally the meshed power and hydrogen network are available.
The biggest cost savings for the hydrogen scenarios derive from the lower need for offshore transmission infrastructure because the offshore wind energy can be integrated through hydrogen.
However, the benefit of transitioning from P2P connections to an offshore network is around 4~bn~\euro{}/a for both electricity-only and hydrogen-inclusive offshore networks.
Interestingly, in the scenarios with hydrogen, the system overall builds more battery storage than in the reference scenario.
This is due to the fact that hydrogen producing countries like Italy export less hydrogen to the NS region, and therefore have more electricity which can be more flexibly used with battery storage.

\begin{figure}[!ht]
    \centering
    \includegraphics[width=.9\textwidth]{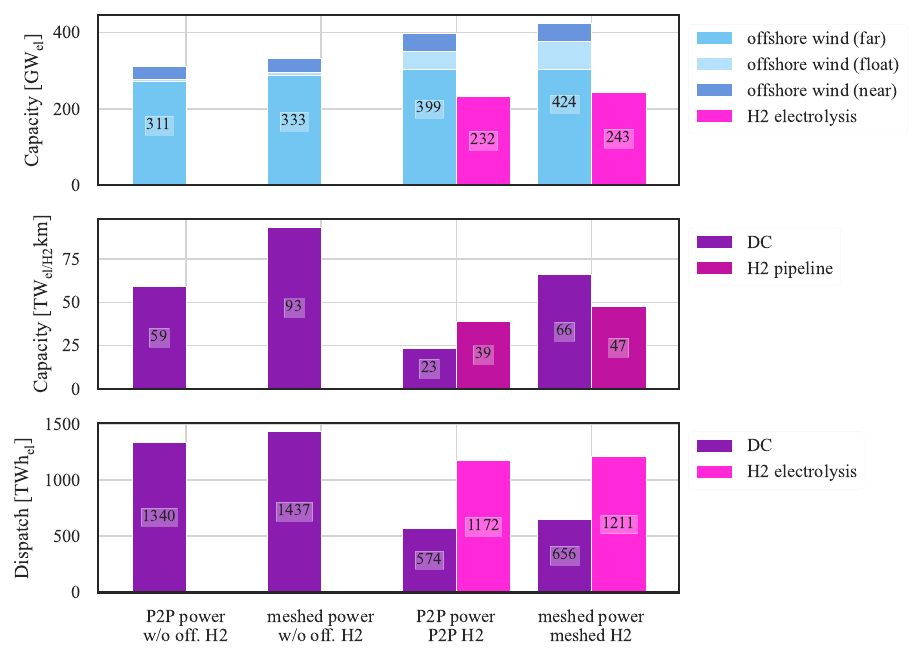}
    \caption{Offshore infrastructure comparison of the different offshore network topologies.  The second row shows the offshore wind and offshore electrolysis capacities. Finally, the last row illustrates the dispatch of offshore wind via HVDC transmission and hydrogen pipelines.}
    \label{fig:ref_infrastructure_comparison}
\end{figure}

Next, we take a closer look at the offshore network infrastructure in Fig.~\ref{fig:ref_infrastructure_comparison}.
The first row displays the offshore wind and offshore electrolysis capacities in the NS.
Offshore wind capacities vary from 311 GW in the scenario with only P2P power connections to 424 GW in the scenario with meshed power and hydrogen network.
Hence, the build-out of both offshore networks allows for an additional 100 GW of cost-effective offshore wind integration.
A comparison between P2P and the meshed network scenarios shows an offshore wind capacity increase of 22 and 25~GW for the power only and power and hydrogen network, respectively.
Interestingly, only for the networks considering P2P and meshed hydrogen networks, floating wind is deployed with capacities of 50~GW and 75~GW.
One reason for this is that the integration of floating wind through hydrogen compared to HVDC transmission is cheaper.
When going from P2P to meshed hydrogen network, the offshore electrolysis capacities increase from 232 to 243~GW.

The second row depicts the offshore HVDC and hydrogen pipeline capacities.
The offshore HVDC capacities increase by approximately 34–43~TWkm when a network is included, irrespective of offshore hydrogen consideration.
However, when we consider hydrogen, the offshore HVDC capacities generally reduce by about 50\%. In the P2P scenario with hydrogen, the model shows a preference for building more hydrogen pipeline than HVDC capacities.

This trend is further highlighted by the last row, which illustrates the proportion of wind energy allocated to HVDC transmission versus hydrogen production.
Approximately one-third of the wind energy takes the electricity route, while two-thirds the hydrogen route.

Analyzing offshore wind curtailment in the NS, we observe from Table~\ref{tab:ref_curtailment} that integrating offshore hydrogen significantly reduces the amount of curtailed offshore wind by half.
The existence of an offshore network also leads to a reduction which is however negligible compared to offshore hydrogen.
Nonetheless, when evaluating curtailment against the total potential offshore wind generation, it is notably low.

\begin{table}[!h]
    \caption{Offshore wind curtailment in North Sea in TWh and in \% of total installed potential.}
    \label{tab:ref_curtailment}
    \begin{tabular}{l|r|r}
    \hline
    & Curtailment [TWh] & Share of total potential [\%] \\
    \hline
    w/o network, w/o H2& 11.38 & 0.84 \\
    with network, w/o H2& 12.79 & 0.88 \\
    w/o network, with H2& 7.61 & 0.43 \\
    with network, with H2& 6.64 & 0.35 \\
    \hline
    \end{tabular}
\end{table}

\subsection{Network topology and capacity allocation}
\label{sec:results_full_network_scenario}

\begin{figure}[!ht]
    \centering
    \includegraphics[width=1\textwidth]{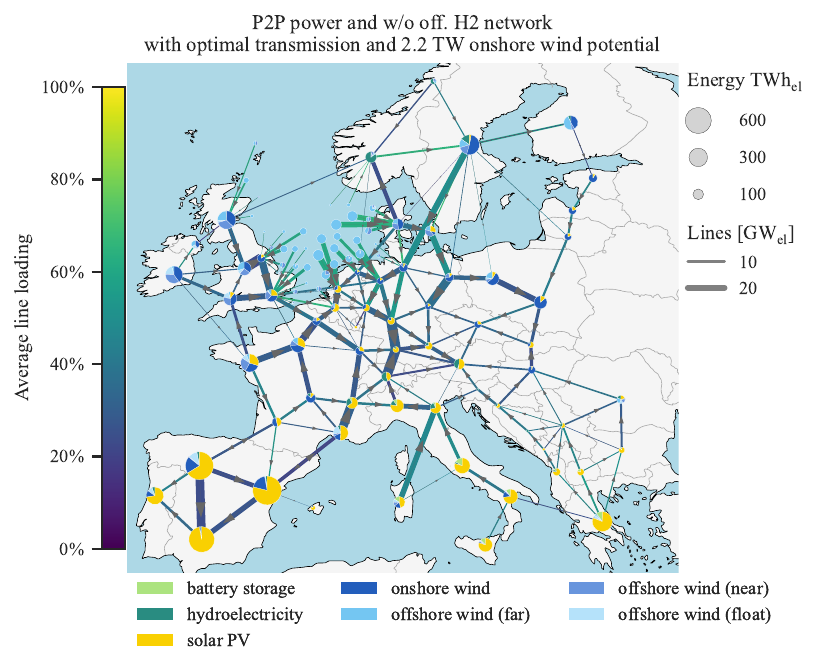}
    \caption{Map of electricity supply and transmission infrastructure for the scenario with point-to-point offshore power network without offshore hydrogen. The color map used for the offshore lines illustrates the average line loading for each specific line.}
    \label{fig:P2P_electricity_map}
\end{figure}

As seen in the previous Section \ref{sec:results_offshore_network_comparison}, the cost benefits are the highest when we allow the system to build both the meshed power and hydrogen network.
Before taking a closer look at the meshed networks, we illustrate the electricity map of the reference scenario only having the P2P power network in Fig.~\ref{fig:P2P_electricity_map}.
The reference scenario mainly integrates fixed bottom wind farms close to shore with good resources, and in countries with high electricity demands or exporting countries like Denmark.
In Norway, only little offshore wind capacities are integrated. 
The line loading of the P2P connections are rather uniform at $\sim$~60\%.
Compared to the average wind capacity factors in the NS of 50\%, this suggests an underdesign in the capacity of P2P connections.
Since we do not have any offshore hydrogen production for the reference case this leads to higher curtailment.
Moreover, the model of the reference scenario expands the TYNDP interconnector project "NorthConnect" between Scotland and Norway.

\begin{figure}[h!]
    \centering
    \includegraphics[width=\textwidth]{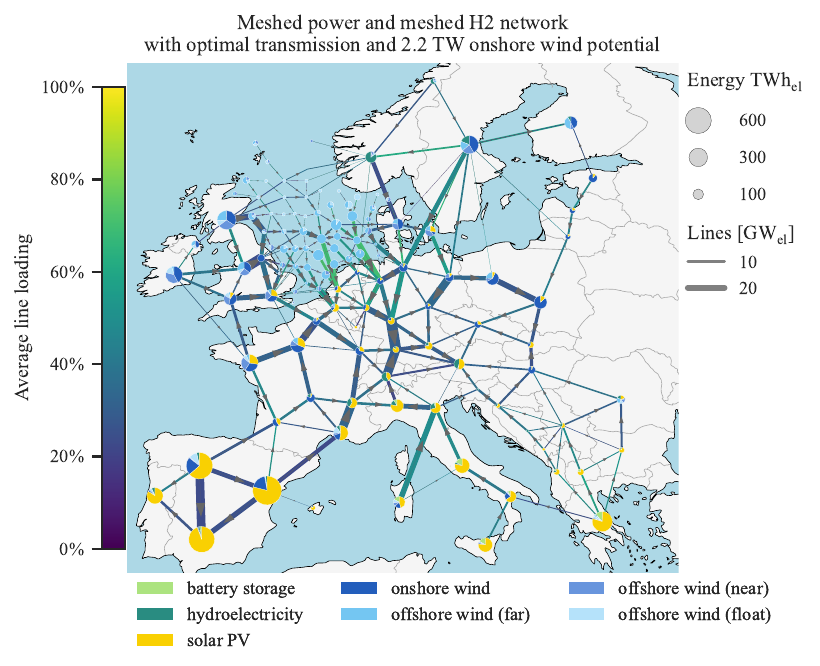}
    \caption{Map of electricity supply and transmission infrastructure for the scenario with offshore power and hydrogen network. The color map used for the offshore lines illustrates the average line loading for each specific line.}
    \label{fig:full_network_electricity_map}
\end{figure}

Fig.~\ref{fig:full_network_electricity_map} shows the electricity network and the supply from different technologies for this scenario.

A key aspect of this scenario is the main corridor, which connects the European mainland with the Great Britain.
Especially from Germany, Belgium and the Netherlands, we observe large interconnection capacity connecting to the offshore grid.
At the same time, the "NorthConnect" interconnector from the reference scenario is substituted by the offshore power network.
Due to the constraint that newly built line capacities cannot exceed 20~GW, many offshore lines are at their maximum capacity.
Furthermore, some lines from this corridor connecting onshore exhibit higher utilization rates compared to other lines with line loading averaging around 70\%.
To have such high line loading, the lines must collect the electricity from multiple surrounding offshore platforms with different wind generations patterns.

Most generation capacity in the NS is far offshore wind with $\sim$300~GW, followed by $\sim$70~GW floating and $\sim$50~GW near offshore wind.
Almost all offshore wind resources for near and far offshore wind in the NS are exhausted.
Certain wind resources are exclusively connected to the offshore grid and not to their home country due to their comparably low energy demand, e.g. in Norway.
Moreover, we observe that there are a couple of locations, mainly with floating wind, which are barely connected to the offshore power network.
The reason for this is the high cost for connecting those offshore wind locations to the offshore network because one would need to install expensive HVDC platforms and cables.
We will later see that those locations are integrated into the offshore hydrogen network.

\begin{figure}[h!]
    \centering
    \includegraphics[width=\textwidth]{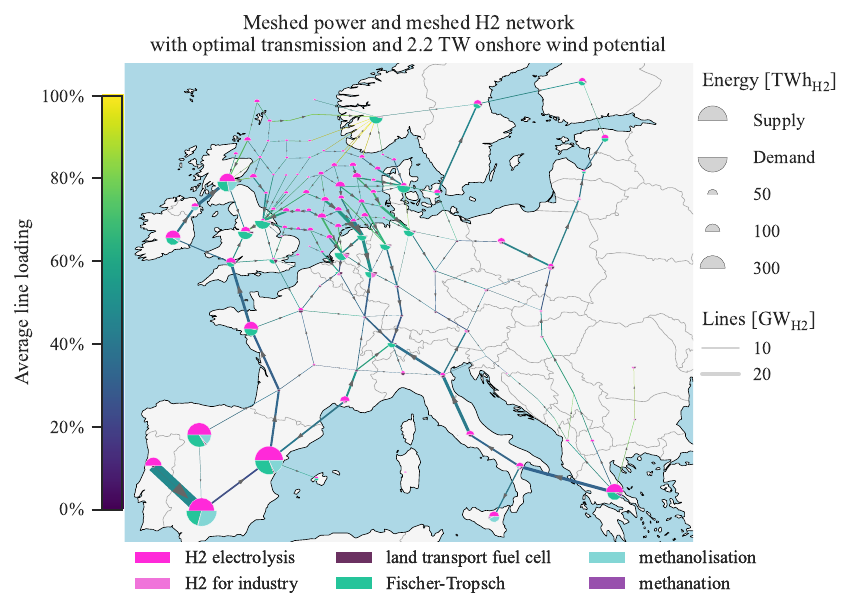}
    \caption{Map of hydrogen supply and demand, and hydrogen infrastructure for the scenario with meshed offshore power and hydrogen network.}
    \label{fig:grid_hydrogen_map}
\end{figure}

Next, we look at the hydrogen network of the scenario with the meshed power and hydrogen network.
In Fig.~\ref{fig:grid_hydrogen_map} we see that all offshore locations are integrated via hydrogen electrolysis and pipelines because they have lower specific connection cost compared to HVDC.
The specific cost difference between HVDC and hydrogen integration are 160~\euro/kW for the conversion and 630~\euro/MW/km for the transport.

The offshore hydrogen network differs recognizably from the electricity network.
The hydrogen interconnections between Great Britain and other NS countries are less pronounced than the HVDC connections.
Bulk hydrogen production for all NS countries except Great Britain occur in the NS.
Since countries like Germany, Belgium and the Netherlands, do not have sufficient onshore hydrogen production to meet their demand, they have to import offshore hydrogen from the NS.
The main hydrogen demand in those countries comes from the direct usage of hydrogen in industry like steel production or from Fischer-Tropsch synthesis to produce kerosene for aviation and naphtha for feedstocks.
To import the offshore hydrogen, the model needs to build hydrogen pipelines.
The hydrogen flow often originates from areas with abundant wind, where the HVDC lines are utilized at rates of 70\%.
This implies that the model just builds enough offshore HVDC capacities that the lines can be operated with such high utilization rates.
For additional export capacities, the model opts for the more cost-effective integration via hydrogen pipelines.
Nonetheless, the fact that the model does not solely build an offshore hydrogen network shows that model prefers to build an HVDC network given that the HVDC lines have a high utilisation rate.

This cost-effectiveness of hydrogen pipelines holds true especially for remote offshore locations, where the model does not build any or insufficient offshore HVDC connections.
Those locations are mainly situated in the North of the NS and are dominated by floating wind installations.
Onshore sites in the NS region exhibit electrolysis capacities only when there is an abundance of renewable energy, such as in the Great Britain, which demonstrates extensive wind resources both onshore, and offshore from the NS and Celtic Sea.
The total onshore electrolysis capacity in the NS reaches to 110~GW with deployments exclusively in Great Britain and Denmark.
Besides the NS, Spain turns out as a primary hydrogen producer in our model.
While the NS in our model produces approximately 25~Mt (830~TWh) of hydrogen, Spain's production reaches similar amounts in the order of 28~Mt (933~TWh).
The map plots for the other scenarios with different offshore network topologies are given in Fig.~\ref{fig:grid_electricity_map} and \ref{fig:P2P_H2_electricity_map}.

\subsection{Effect of Onshore Wind Potential and Onshore Transmission Expansion}
\label{sec:results_sensitivity}

\begin{figure}[!ht]
    \centering
    \includegraphics[width=.7\textwidth]{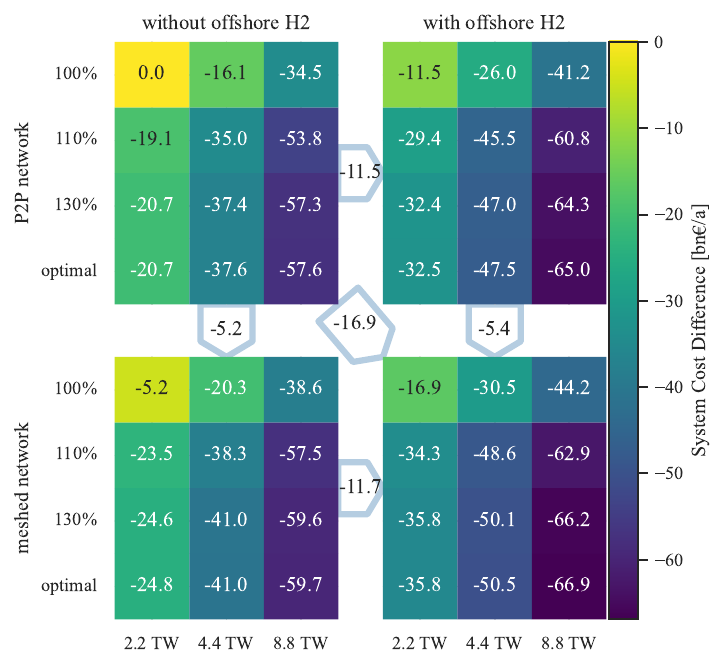}
    \caption{Sensitivity of total system costs to onshore wind potential and onshore transmission expansion for the different offshore network topologies.}
    \label{fig:total_cost_sensitivity}
\end{figure}

In this section, we explore the impact of onshore wind potentials and onshore transmission expansion on different offshore network topologies.
We change the onshore wind potential within a range of 2.2 to 8.8~TW and adjust onshore transmission from a baseline of 100\% to the optimal expansion.
Fig.~\ref{fig:total_cost_sensitivity} illustrates the effects of varying the parameters on system costs.
The figure depicts quadrants, each representing one of the four network topologies of our study i.e. two without network and two with network where we vary the presence of offshore hydrogen production.
For example in the first quadrant, we depict only the scenarios with only P2P power network and without hydrogen.
To measure the system cost effects for different network topologies, we take the most constrained case of the first quadrant as the base case with an onshore wind potential of 2.2~TW and 100\% transmission.
For the other combinations of onshore wind and transmission in each quadrant, we then calculate the cost differences relative to the base case.
It is important to note that negative values in this context indicate a reduction in costs compared to the base case.

Consistent with the results from the previous section \ref{sec:results_offshore_network_comparison}, we observe that the meshed offshore power and hydrogen network have the highest cost benefits compared to the other topologies.
This hold true even if we vary the onshore wind and transmission limits.
However, when comparing the upper left and lower right quadrant, we observe that the benefits of the meshed network becomes less starting at 16.9~bn\euro/a reducing to 9.3~bn\euro/a when we relax the onshore wind from 2.2 to 8.8~TW and transmission constraint from 100\% to optimal.

\begin{figure}[h!]
    \centering
    \includegraphics[width=.7\textwidth]{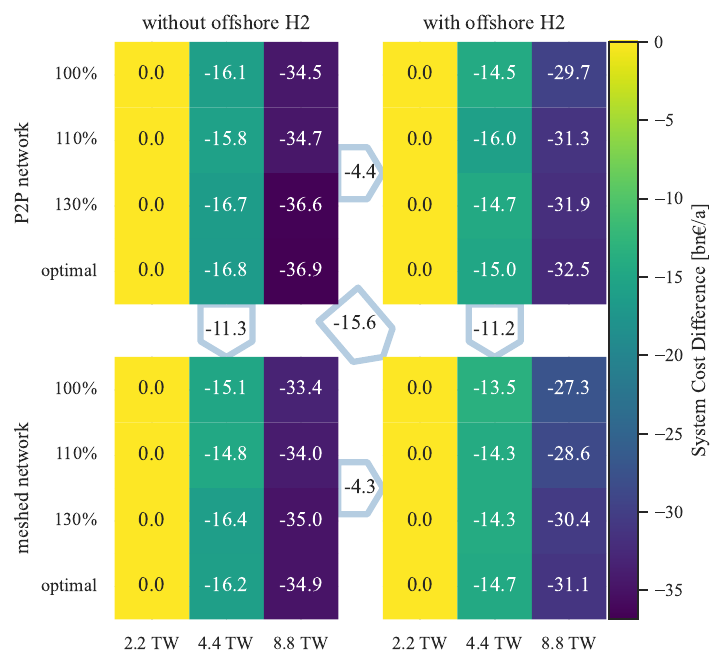}
    \caption{Total system cost sensitivity depending on onshore wind potential. Each column within a quadrant is compared to the left column with 2.2~TW onshore wind potential.}
    \label{fig:sensitivity_wind_comparison}
\end{figure}

To derive how the benefits of the topologies change with onshore wind potential, we can take a look at Fig.~\ref{fig:sensitivity_wind_comparison}.
In this figure, we calculate the differences relative to the first column in each quadrant.
The arrows between the quadrant show the average cost difference between the first column of the different quadrants.
Although it initially appears that the cost benefit is linear when increasing the potential from 2.2 to 8.8~TW, this is not the case.
The cost advantages are similar around 14–18~bn\euro/a when increasing from 2.2 to 4.4 or from 4.4 to 8.8~TW, even though the potentials double and quadruple, respectively.
Furthermore, in the left quadrants, the cost benefits from increasing onshore wind potential are bigger than in the right quadrants.
One reason for this is that offshore hydrogen can serve parts of the onshore hydrogen demand which would otherwise be produces by onshore wind.

\begin{figure}[h!]
    \centering
    \includegraphics[width=.7\textwidth]{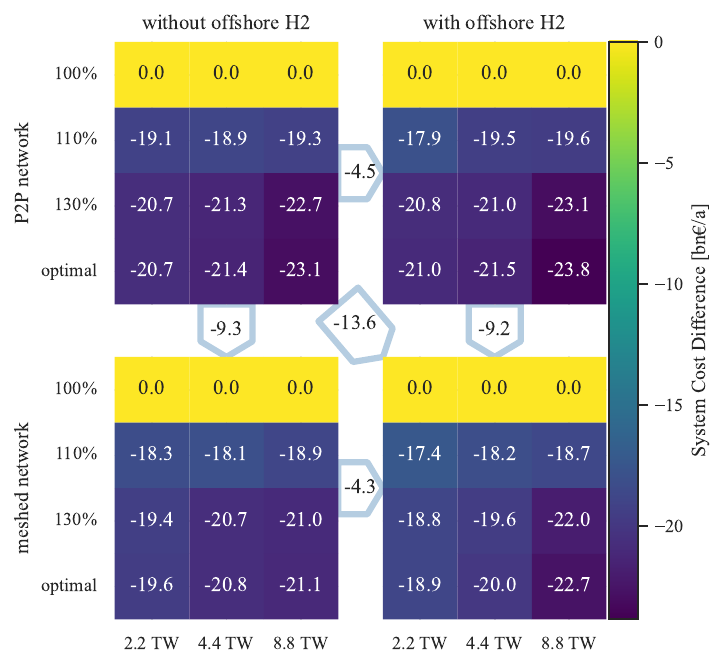}
    \caption{Total system cost sensitivity depending on onshore transmission expansion. Each row within a quadrant is compared to the top row with 100\% onshore transmission limit.}
    \label{fig:sensitivity_tranmission_comparison}
\end{figure}

Next, we turn to the impact of onshore transmission expansion on system cost in Fig.~\ref{fig:sensitivity_tranmission_comparison}.
Comparing the rows demonstrates that the cost reduction effect is not linear; the most significant benefits of 17–19~bn.~\euro{}/a are observed when transitioning from 100\% to 110\% onshore transmission expansion, regardless of the network topology.
From 110\% to optimal onshore transmission, the cost benefits reduce to $\sim$2-4~bn.~\euro{}/a. 
Hence, similar to the onshore wind potential, the cost benefits are larger for the first transmission increase to 110\%.
Moreover, the meshed networks benefit 1–2~bn\euro/a less from the additional onshore transmission expansion compared to the P2P networks.
The reason for this is that the offshore power network can use alternative routes to transport electricity from offshore wind farms onshore avoiding congested onshore transmission lines.

In summary, the sensitivity analysis shows that the cost benefits of an offshore power and hydrogen network are consistently the highest, regardless of the onshore wind potential and transmission limits.
Therefore, the results of Section \ref{sec:results_offshore_network_comparison} are valid for all considered onshore wind potentials and transmission limits.
For both onshore wind and transmission, the cost benefits are the highest for the first relaxation step from 2.2~TW to 4.4~TW and from 100\% to 110\%, respectively.
Furthermore, with limited onshore transmission, a meshed offshore power network better compensates, while for limited onshore wind potentials, offshore hydrogen production is more advantageous.
\section{Discussion}
\label{sec:Discussion}

In our analysis we observe a strong impact of offshore hydrogen production on the total system cost.
This is contrary to the studies of Gea-Bermúdez et al.~\cite{gea-bermudezGoingOffshoreNot2023} which find that hydrogen production in the North Sea plays a limited role and most production takes place onshore.
However, our analysis finds that in the North Sea region the majority of hydrogen production happens offshore. 
One reason for this could be that our endogenously modelled demand for hydrogen in the North Sea countries is twice as high compared to~\cite{gea-bermudezGoingOffshoreNot2023} with 2000~TWh, since it includes industrial feedstocks and fuels for aviation and shipping.
Furthermore, they do not limit the onshore wind potential, thereby allowing a larger fraction of onshore hydrogen production.

Lüth et al.~\cite{luthHowConnectEnergy2023} see a benefit of having an offshore power network combined with offshore electrolysis.
Similar to our study, offshore electrolysis is found relevant for remote sites because of the high connection cost with submarine HVDC cables.
Similar to Gea-Bermúdez et al.~\cite{gea-bermudezGoingOffshoreNot2023}, they conclude that onshore electrolysis is more important in the North Sea region than offshore electrolysis.
Furthermore, Lüth et al.~discover that onshore transmission congestion drives offshore electrolysis.
However, we observe rather stable offshore electrolysis capacities varying by 1–3~GW even if we have more onshore congestion, see Fig.~\ref{fig:sensitivity_electrolysis_comparison}.
One reason for this is that our model also builds offshore electrolysis when we have optimal onshore transmission capacities, due to the lower cost of connecting offshore wind farms through hydrogen pipelines instead of HVDC cables.

Consistent with our findings, Martínez-Gordón et al.~\cite{martinez-gordonBenefitsIntegratedPower2022} identify a 3-times larger cost benefit when allowing offshore hydrogen production in a meshed offshore network compared to only having a meshed power network.
This might be because, like us, they model a higher endogenous hydrogen demand for whole Europe, not just the North Sea region. 
They also find that onshore wind potentials have a strong impact on the cost-effectiveness of the offshore networks.
Similar to the review in~\cite{gorensteindedeccaReviewNorthSeas2016}, we can conclude that meshed offshore power networks show only a marginal benefit compared to P2P connections.

\section{Limitations}
\label{sec:limitations}
This study has several limitations.
First, the power network in our analysis employs a meshed offshore DC topology, but it does not incorporate the Kirchhoff voltage law for DC lines.
This means that from a physical standpoint, our model does not accurately represent a meshed power grid, as the lines are modeled as simple transport model.

Moreover, all offshore platforms in the North Sea have HVDC connections, and we do not consider AC connections.
The wind farms are connected to the offshore platforms which means that the wind farms cannot directly connect onshore with AC cables.
The offshore platform locations are not located according to the wind farm locations, but are located such that we have an even distribution of platforms in the North Sea.
This could mean that some platform locations could be suboptimal and could be in areas where there are few wind farms.

Additionally, we do not take into account some ecological constraints for the deployment of offshore infrastructure.
For example, we neglect impacts on marine life and habitat deriving from the installation or operation of offshore infrastructure, such as wind turbines and platforms.
Furthermore, for submarine cables and pipelines, we do not consider realistic corridors where crossing of nature reserves or sand extraction areas is prohibited.

Lastly, the model employed in this study is linear and, as such, cannot make discrete investment decisions.
Those discrete decisions variables are important to more accurately reflect real world investment decisions like subsea cables and pipelines as well as offshore platforms.
However, by aggregating at large scale, the discreteness plays less of a role, and we benefit computationally by linearizing.

\section{Conclusion}
\label{sec:conclusion}
In this paper, we investigate the benefits of having an offshore power and hydrogen network in the North Sea and what cost-effective power and hydrogen network topologies might look like.
For this, we minimize the total system cost using an extensive sector-coupled energy system optimization model with 64 onshore regions for Europe and 66 offshore regions within the North Sea.
The model depicts a carbon-neutral scenario for 2050 with a 3-hourly resolution.
With the especially high spatial resolution in the North Sea, we want to let the model endogenously decide which infrastructure it builds to cost-efficiently integrate offshore wind resources.
For our analysis, we evaluate four different offshore network scenarios.
The different scenarios consist of an offshore point-to-point~(P2P) power network without offshore hydrogen, an offshore meshed power network without offshore hydrogen, an offshore P2P power and hydrogen network and finally an offshore meshed power and hydrogen network.

The scenario with only P2P power network costs around 800~bn\euro/a.
When we additionally allow meshed offshore power network connections, the total system cost reduces by 4~bn~\euro{}/a (0.5\% of total cost) and builds 22~GW more (330~GW in total) offshore wind capacity in the North Sea compared to the P2P power network.
The scenario with both offshore power and hydrogen networks has the biggest cost benefit of 15~bn~\euro{}/a (2\% of total cost) compared to the scenario with P2P power network without offshore hydrogen production.
Furthermore, we observe 110~GW more (420~GW in total) offshore wind capacity in the North Sea and a total offshore electrolysis capacity of 240~GW compared to 1080~GW onshore electrolysis of which 60\% is on the Iberian Peninsula.
Only in the scenarios with offshore hydrogen production do we observe a notable deployment of floating offshore wind with 50~GW and 75~GW with P2P and meshed hydrogen network, respectively.

In a sensitivity analysis on the impact of increasing onshore wind potential and onshore transmission expansion, we observe that onshore wind potential has the biggest lever on total system cost for all considered topologies.
However, having more offshore network options like the meshed power and hydrogen network reduce the welfare loss of both limited onshore wind potential or onshore transmission expansion by around 7 and 2~bn\euro{}/a, respectively.

We conclude that a meshed offshore power network in the North Sea is beneficial for the European energy system.
However, having offshore hydrogen production and transport infrastructure is even more advantageous.
Our results indicate that floating offshore wind in the North Sea is only deployed when we have the option to integrate it with offshore hydrogen production. 

\newpage
\appendix
\counterwithin{figure}{section}
\counterwithin{table}{section}
\section{}

\subsection{Study Cases Parameters}\label{tab:study-case-parameters}

\begin{table}[!ht]
    \centering
    \caption{Countries considered in the model.}
    \label{tab:model_countries}
    \begin{tabular}{ llll }
    Albania & Austria & Bosnia and Herzegovina & Belgium \\
    Bulgaria & Switzerland & Czech Republic & Germany \\
    Denmark & Estonia & Spain & Finland \\
    France & United Kingdom & Greece & Croatia \\
    Hungary & Ireland & Italy & Lithuania \\
    Luxembourg & Latvia & Montenegro & North Macedonia \\
    Netherlands & Norway & Poland & Portugal \\
    Romania & Serbia & Sweden & Slovenia \\
    Slovakia &  &  &  \\
    \end{tabular}
\end{table}

\begin{table}[h!]
    \centering
    \caption{Cost assumptions for selected technologies. Cost for fixed bottom installations in North Sea we apply the cost model shown in Fig.~\ref{fig:offwind_cost}. For more cost assumptions and sources refer to \href{https://github.com/p-glaum/technology-data/blob/master/outputs/costs_2030.csv}{Technology Database}.}
    \label{tab:cost_assumptions}
    \begin{tabular}{ lrr }
        Technology&Value&Unit\\
        \hline
        HVDC overhead cable& 430&\euro{}/MW/km\\
        HVDC submarine cable& 960&\euro{}/MW/km\\
        Offshore HVDC platform& 600&\euro{}/kW\\
        Hydrogen pipeline&226&\euro{}/MW/km\\
        Hydrogen submarine pipeline& 329&\euro{}/MW/km\\
        Electrolysis onshore& 400&\euro{}/kW\\
        Electrolysis offshore& 440&\euro{}/kW\\
        Floating offshore wind& 2000&\euro{}/kW\\
        Nearshore offshore wind (constant)& 1250&\euro{}/kW\\
        Far offshore wind (constant)& 1600&\euro{}/kW\\
    \end{tabular}
\end{table}

\newpage
\subsection{Reference Scenario Plots}
\label{sec:scenario-plots}
\begin{figure}[h!]
    \centering
    \includegraphics[width=1\textwidth]{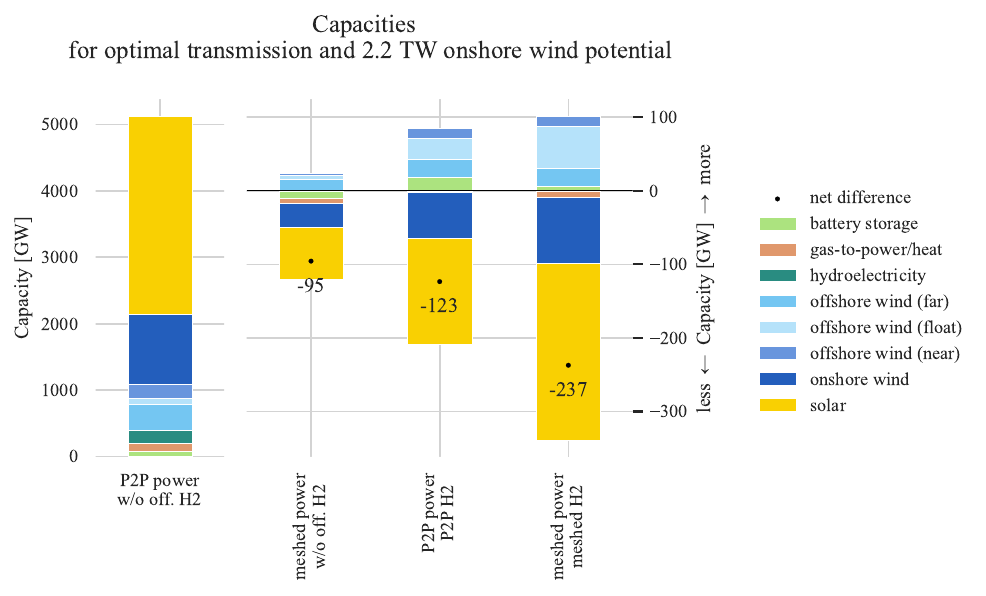}
    \caption{Capacity comparison of the different offshore network topologies. The first column shows the reference case. In the other columns we calculate the capacity differences compared to the reference case.}
    \label{fig:ref_capacity_comparison}
\end{figure}

\begin{figure}[h!]
    \centering
    \includegraphics[width=.7\textwidth]{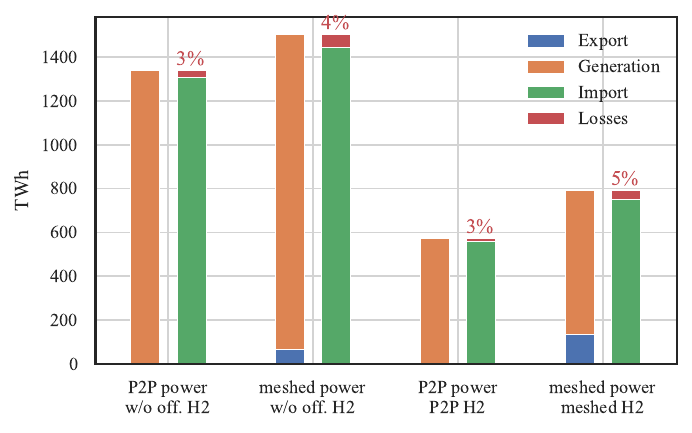}
    \caption{Import and export electricity transmission comparison of the different offshore network topologies.}
    \label{fig:ref_tranmission_comparison}
\end{figure}

\begin{figure}[h!]
    \centering
    \includegraphics[width=1\textwidth]{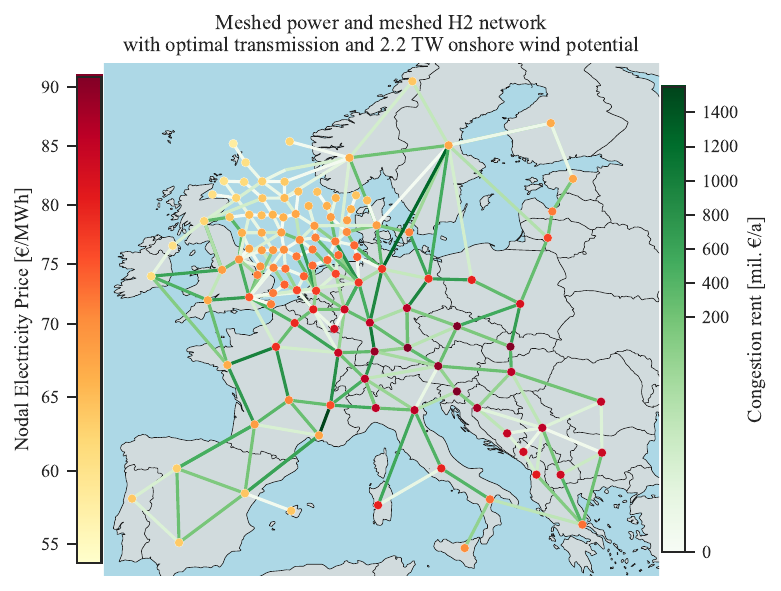}
    \caption{Nodal electricity prices and congestion rent for scenario with meshed power and hydrogen network, optimal onshore transmission and 2.2~TW onshore wind potential. The average nodal price in the North Sea is 69~\euro/MWh.}
    \label{fig:price_map_100_transmission}
\end{figure}

\begin{figure}[h!]
    \centering
    \includegraphics[width=1\textwidth]{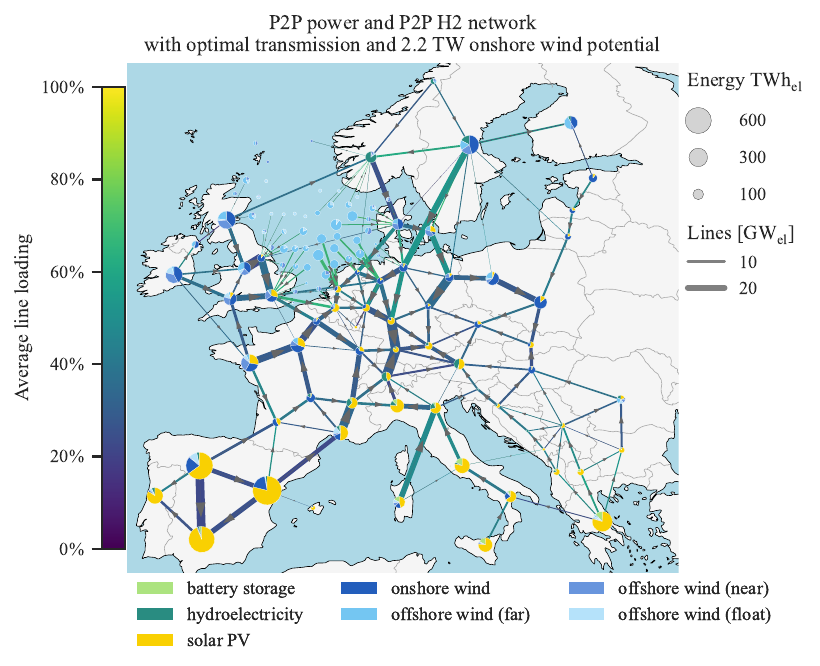}
    \caption{Map of electricity supply and transmission infrastructure for the scenario with point-to-point offshore power and hydrogen network. The color map used for the offshore lines illustrates the average line loading for each specific line.}
    \label{fig:P2P_H2_electricity_map}
\end{figure}

\begin{figure}[h!]
    \centering
    \includegraphics[width=1\textwidth]{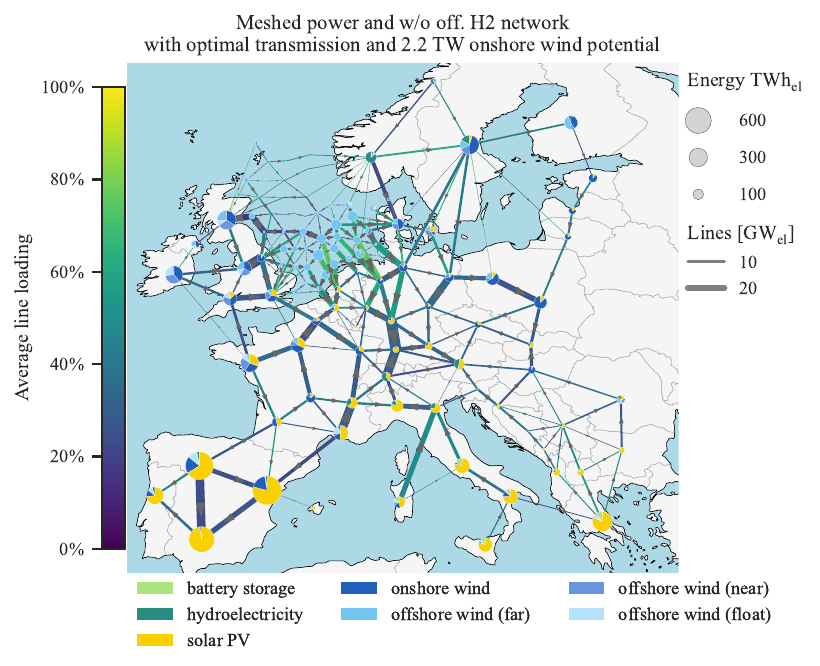}
    \caption{Map of electricity supply and transmission infrastructure for the scenario with meshed offshore network without offshore hydrogen. The color map used for the offshore lines illustrates the average line loading for each specific line.}
    \label{fig:grid_electricity_map}
\end{figure}

\begin{figure}[h!]
    \centering
    \includegraphics[width=1\textwidth]{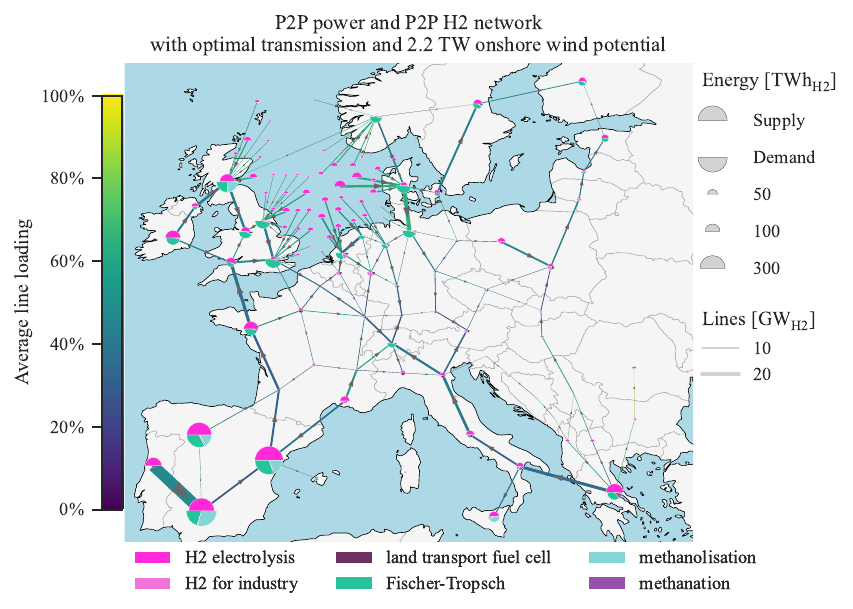}
    \caption{Map of hydrogen supply and demand, and hydrogen infrastructure for the scenario with offshore power and hydrogen P2P network.}
    \label{fig:P2P_H2_hydrogen_map}
\end{figure}

\FloatBarrier
\subsection{Sensitivity}
\label{sec:sensitivity}

\begin{figure}[h!]
    \centering
    \includegraphics[width=.7\textwidth]{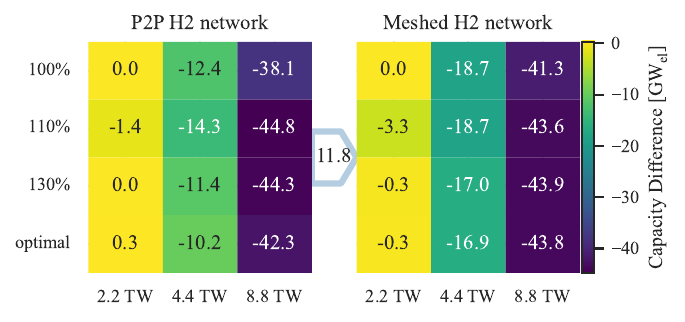}
    \caption{Sensitivity of offshore electrolysis capacities in the North Sea to onshore wind potential and onshore transmission expansion. Each field within a quadrant is compared to the upper left field.}
    \label{fig:sensitivity_electrolysis_comparison}
\end{figure}

\begin{figure}[h!]
    \centering
    \includegraphics[width=.7\textwidth]{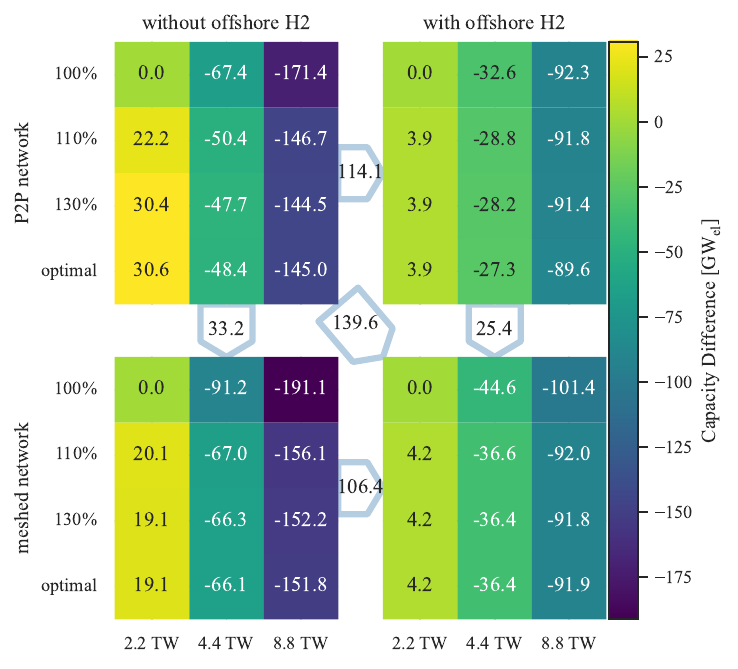}
    \caption{Sensitivity of offshore wind capacities in the North Sea to onshore wind potential and onshore transmission expansion for the different offshore network topologies.}
    \label{fig:offshore_wind_capacity_sensitivity}
\end{figure}

\clearpage
\bibliographystyle{elsarticle-num}
\bibliography{references} %

\end{document}